\documentclass[12pt]{article}
\usepackage{amssymb,graphicx}
\usepackage{epstopdf}
\usepackage{amsmath,amsfonts}
\usepackage{epsfig}

\def\d{\partial}
\def\l{\left(}
\def\r{\right)}

\newcommand{\be}{\begin{equation}}
\newcommand{\ee}{\end{equation}}
\newcommand{\bea}{\begin{eqnarray}}
\newcommand{\eea}{\end{eqnarray}}
\newcommand{\bg}{\begin{gather}}
\newcommand{\eg}{\end{gather}}
\newcommand{\bseq}{\begin{subequations}}
\newcommand{\eseq}{\end{subequations}}

\begin{document}
\begin{flushright}
INR-TH-2011-14
\end{flushright}
\vspace{10pt}
\begin{center}
  {\LARGE \bf Dynamical vs spectator models
of (pseudo-)conformal Universe} \\
\vspace{30pt}
M.~Libanov and  V.~Rubakov\\
\vspace{15pt}
\textit{
Institute for Nuclear Research of
         the Russian Academy of Sciences,\\  60th October Anniversary
  Prospect, 7a, 117312 Moscow, Russia}\\
\vspace{5pt}
\textit{
Physics Department, Moscow State University,\\ Vorobjevy Gory,
119991, Moscow, Russia
}

\vspace{5pt}
    \end{center}
    \vspace{5pt}

\begin{abstract}
We discuss two versions of the conformal scenario for generating scalar
cosmological perturbations: a spectator version with a scalar field
conformally coupled to gravity and carrying negligible energy density, and
a dynamical version with a scalar field minimally coupled to gravity and
dominating the cosmological evolution. By making use of the Newtonian
gauge, we show that (i) no UV strong coupling scale is generated below
$M_{Pl}$ due to mixing with metric perturbations in the dynamical
scenario, and (ii) the dynamical and spectator models yield identical
results to the leading {\it non-linear} order. We argue that these
results, which include potentially observable effects like statistical
anisotropy and non-Gaussianity, are characteristic of the entire class of
conformal models. As an example, we reproduce, within the dynamical
scenario and working in comoving gauge, our earlier result on the
statistical anisotropy, which was originally obtained within the spectator
approach.
\end{abstract}
\section{Introduction}

Recently, an idea of attributing the flatness of the scalar spectrum of
the primordial cosmological perturbations to approximate conformal
symmetry, rather than approximate de~Sitter symmetry of inflationary
theory, has attracted some interest~\cite{vr1,Cr,HK} (see
Ref.~\cite{Antoniadis:1996dj} for earlier work). In the simplest version
of the conformal scenario, one assumes that the gravity effects are
totally negligible and considers a theory of two fields $\phi$ and
$\theta$ of conformal weights 1 and 0, respectively\footnote{The conformal
weight of $\phi$ may be different from 1, it is important only that this
weight is non-zero~\cite{HK}.}, with the Lagrangian (mostly negative
signature)
\be
L_{\phi, \theta} = L_{\phi} + \frac{1}{2} \phi^2 \l \d_\mu \theta \r^2
\label{1*}
\ee
in effectively Minkowski space-time. In other words, this version
assumes that the field $\phi$ (as well as $\theta$) is a spectator whose
dynamics does not affect the space-time metric. An example is a theory of
complex scalar field conformally coupled to gravity, whose energy density
is negligible compared to the total energy density in the Universe, with
$\phi$ and $\theta$ identified with the modulus and phase of that complex
field~\cite{vr1}.

The background field $\phi_c$ is assumed to be spatially homogeneous and
evolving non-trivially; then conformal invariance implies that
\be
\phi_c (t) = - \frac{\sqrt{2}}{\sqrt{\lambda}t} \;, \;\;\;\;\;\; t <0
\; ,
\label{1**}
\ee
where $\lambda $ is a dimensionless constant and one assumes that
$\lambda \ll 1$ for canonically normalized $\phi$. The notations here are
chosen in such a way that $\phi_c(t)$ is a solution in a theory with
negative quartic potential,
\be
L_\phi = \frac{1}{2} \l \d_\mu \phi \r^2 - V(\phi) \; , \;\;\;\;\;\;\;
V(\phi) = - \frac{\lambda}{4} \phi^4 \; .
\label{2*}
\ee
In this background, the field $\theta$ starts off in the WKB regime
which holds as long as $k|t| \gg 1$ (short wavelengths), where $k$ is the
spatial momentum of a $\theta$-mode. One naturally considers the initial
vacuum state. Fluctuations of $\theta$ freeze out at late times, when
$k|t| \ll 1$ (large wavelength regime). These fluctuations have flat power
spectrum and are thought of as precursors of the adiabatic perturbations.
The conversion of $\theta$-perturbations into adiabatic ones occurs at
much later stage via, e.g., curvaton~\cite{Linde:1996gt} or modulated
decay~\cite{Dvali:2003em} mechanism.

A peculiar feature of this scenario is the existence of the perturbations
of the field $\phi$ itself. In spectator models, these perturbations  have
{\it red} power spectrum in the large wavelength regime~\cite{vr1,HK}.
Interaction of the field $\theta$ with perturbations of $\phi$ leads to
potentially observable effects, such as statistical anisotropy and
specific forms of
non-Gaussianity~\cite{Libanov:2010nk,Libanov:2010ci,Libanov:2011hh,Mironov-2}.
These effects are quite generic, since both the properties of
$\phi$-perturbations and their interaction with $\theta$ are dictated by
conformal invariance. The latter point is discussed, within the spectator
approach, in Refs.~\cite{HK,Mironov-2}.

Instead of considering the spectator version of the conformal scenario, it
is of interest to study dynamical (pseudo-)conformal models, i.e., treat
the fields $\phi$, $\theta$ as the only  matter fields relevant at the
early epoch. In this class of models, the rolling field $\phi_c (t)$
determines the homogeneous evolution of the Universe, while the
perturbations of $\phi$ come together with metric perturbations. One model
of this sort has been proposed by Hinterbichler and Khoury~\cite{HK}, who
considered a theory with the action
\be
S = - \frac{M_{Pl}^2}{2} \int~d^4x~\sqrt{-g} R + \int~d^4x~\sqrt{-g}
L_{\phi \theta} \; ,
\label{jun28-1}
\ee
where the scalar Lagrangian is given by Eqs.~\eqref{1*}, \eqref{2*}.
In the Hinterbichler--Khoury model, the rolling scalar field \eqref{1**}
has super-stiff equation of state, $p_c \gg \rho_c > 0$, and the Universe
 contracts. So, this model is in a certain sense a reincarnation of the
ekpyrosis scenario~\cite{ekpyro-i}. Another model is Galilean
Genesis~\cite{Cr}, in which the field $\phi$ is conformal
Galileon~\cite{Nicolis:2008in}. In the latter model, pressure is negative
and violates null energy condition, space-time is initially Minkowskian,
the energy density increases in time, expansion speeds up, and eventually
transition to the hot epoch occurs in some way (see
 Ref.~\cite{Levasseur:2011mw} for the discussion of the last,
``defrosting'' stage). Clearly, both of these models are interesting
alternatives to the inflationary scenario.

In dynamical conformal models, the properties of the perturbations
$\varphi = \phi - \phi_c$ and associated metric perturbations are somewhat
subtle. While the power spectrum of $\varphi$ is red in the absence of
gravity (in the large wavelength regime), the power spectrum of the
curvature perturbations $\zeta$ is {\it blue} in dynamical
models~\cite{Cr,HK} (see also Ref.~\cite{Gao:2011mz}). This feature has
lead Hinterbichler and Khoury~\cite{HK} to argue, on the basis of
power-counting, that the theory with the scalar Lagrangian \eqref{2*} has
UV strong coupling scale
\be
\Lambda^{(1)} = \frac{\phi_c^3}{M_{Pl}^2} \; ,
\label{4*}
\ee
which is time-dependent and small at early times. They also argued
that adding the field $\theta$ with the Lagrangian written in \eqref{1*}
yields another UV strong coupling scale
\be
\Lambda^{(2)} = \lambda^{1/4} \phi_c \; .
\label{4**}
\ee
With so low strong coupling scales, self-consistency of the model
would imply extremely strong constraints on the self-coupling parameter
$\lambda$~\cite{HK}.

Another side of the subtlety with $\varphi$-perturbations is that their
mixing with metric is apparently important for all scales, at least in the
large wavelength regime. So, one may doubt that the results of
Refs.~\cite{Libanov:2010nk,Libanov:2010ci,Libanov:2011hh,Mironov-2},
obtained within the spectator approach, are valid in dynamical conformal
models as well.

In this note we intend to clarify these issues, making use of the
Hinterbichler--Khoury model as an example. Concerning strong coupling in
UV, we recall that naive power counting may or may not give correct
results depending on the gauge choice. A famous example is given by
non-Abelian gauge theories with the Higgs mechanism, where power counting
in the unitary gauge suggests the UV strong coupling scale of the order of
the Higgs expectation value, while power counting in $R_\xi$ gauge shows
that there is no strong coupling in UV at all\footnote{Modulo the Landau
pole, which is another story.}. In the Hinterbichler--Khoury model, as
well as in Galilean Genesis, the energy density is small at relevant
times, and space-time is almost flat. It is well known that in such a
situation it is appropriate to make use of the Newtonian gauge. We employ
the Newtonian gauge in this note and show that metric perturbations are in
fact small\footnote{This has been pointed out in Ref.~\cite{Cr} in the
context of Galilean Genesis.}, non-linear gravitational effects are
suppressed, and no UV strong coupling scale is generated below $M_{Pl}$.
Hence, the UV scales \eqref{4*} and \eqref{4**} are actually not there. As
a consequence, one does not have to impose strong constraints on the
self-coupling $\lambda$.

Likewise, in the Newtonian gauge the main non-linear effects are due to
the interaction of $\theta$ with perturbations $\varphi$, while the
effects coming from the interaction of $\theta$ with metric perturbations
are small. This shows that the analysis of
Refs.~\cite{Libanov:2010nk,Libanov:2010ci,Libanov:2011hh,Mironov-2} does
apply to dynamical conformal models. As an illustration of this general
conclusion, we rederive one of the results of Ref.~\cite{Libanov:2010nk}
(statistical anisotropy) within the dynamical model and in the gauge
$\varphi=0$, where the gravitational potential $\Psi$ coincides with
$\zeta$ and has blue power spectrum. This derivation also sheds some light
on the cancellation of infrared effects, which was somewhat surprising in
the spectator approach~\cite{Libanov:2010nk}.

This paper is organized as follows. We introduce the Hinterbichler--Khoury
model in Section~\ref{sec:model}. In Section~\ref{sec:scales} we discuss
the strong coupling issue. The leading non-linear effects are considered
in Section~\ref{sec:non-linear}. We conclude in
Section~\ref{sec:conclusion}.

\section{The model}
\label{sec:model}

We consider the model with the action \eqref{jun28-1} where the scalar
Lagrangian is given by \eqref{1*}, \eqref{2*}. We first recall the
properties of the homogeneous, spatially flat background at early times.
As $t \to -\infty$, the gravitational effects on the background field
$\phi_c$ are negligible, and $\phi_c$ rolls down negative quartic
potential according to \eqref{1**}. The pressure is
\[
p_c = \dot{\phi_c}^2 = \frac{2}{\lambda t^4} \; ,
\]
while the energy density is small, $\rho_c \ll p_c$. Upon integrating
the Raychaudhury equation one finds the Hubble parameter
\be
H = \frac{1}{3\lambda t^3 M_{Pl}^2} \; .
\label{jun28-2}
\ee
The Universe contracts, and matter in it has super-stiff equation of
state. This regime persists as long as\footnote{As $\phi_c$ approaches
$M_{Pl}$, the effect of the cosmological expansion on the evolution of
$\phi_c$ becomes important ($H\dot{\phi}_c$ becomes comparable to
$\ddot{\phi}_c$), energy density catches up with  pressure, and the regime
\eqref{1*}, \eqref{jun28-2} terminates.}
\be
\lambda t^2 M_{Pl}^2 = \frac{2 M_{Pl}^2}{\phi_c^2} \gg 1 \; .
\label{jun28-3}
\ee
In this paper we consider exclusively early times, when the inequality
\eqref{jun28-3} holds. It is worth noting that we can consistently set the
scale factor equal to 1 wherever it enters without time derivative(s), and
identify conformal time and cosmic time.

We are primarily interested in perturbations $\varphi = \phi - \phi_c$ and
associated metric perturbations. To simplify the formulas below, we
partially fix the gauge by setting the longitudinal component of spatial
metric perturbation equal to zero, so that
\be
ds^2 = (1+2\Phi) dt^2 + 2 \d_i U dt dx^i - (1+2\Psi) \delta_{ij} dx^i
dx^j \; .
\label{jun29-1}
\ee
We recall that in a general theory of one scalar field with canonical
kinetic term, the gauge invariant combination (prime denotes the
derivative with respect to conformal time)
\be
\frac{v}{a} = \varphi - \frac{a\phi_c^\prime}{a^\prime} \Psi
\label{jun28-6}
\ee
is canonically normalized and obeys the equation
\be
v^{\prime \prime} - \Delta v - \frac{z^{\prime \prime}}{z} v = 0\;,
\label{jun28-5}
\ee
where
\[
z= \frac{a^2 \phi_c^\prime}{a^\prime} \; .
\]
In the case at hand $a=1$ and $z= \dot{\phi}_c/H$, so the last term in
the left hand side of Eq.~\eqref{jun28-5} is negligible (it is of order
$(\lambda t^4 M_{Pl}^2)^{-1} \cdot v$, see Ref.~\cite{HK}). Hence, the
field $v$ is a free scalar field in effectively Minkowski space-time. The
variable $\zeta$, the curvature perturbation at comoving hypersurfaces, is
related to $v$ by
\[
\zeta = - \frac{a^\prime}{a^2 \phi_c^\prime} \cdot v = -
\frac{H}{\dot{\phi}_c} \cdot v\; .
\]
Both $v$ and $\zeta$ have blue power spectrum~\cite{Cr,HK}.

\section{Strong coupling scales and absence thereof}
\label{sec:scales}

\subsection{Spurious scales $\Lambda^{(1)}$ and $\Lambda^{(2)}$: gauge
$\varphi=0$}

To see the spurious appearance of the UV scales \eqref{4*} and
\eqref{4**}, let us choose the gauge $\varphi = 0$. In this gauge, the
gravitational potential $\Psi$ coincides with $\zeta$, so our discussion
parallels that of Hinterbichler and Khoury~\cite{HK}, who worked with the
gauge-invariant variable $\zeta$. Let us consider short wavelength modes,
$k|t| \gg 1$. Then in the gauge $\varphi = 0$ one has from \eqref{jun28-6}
\[
\Psi \sim \frac{1}{\sqrt{\lambda} M_{Pl}^2 t} v \; .
\]
The constraint equations in this gauge are
\begin{align}
\Delta U &=  - 3 \Psi^\prime + \frac{a}{a^\prime} \Delta \Psi +
\frac{a^3}{a^{\prime }} \frac{\rho_c - p_c}{2M_{Pl}^2}\Phi = - 3
\dot{\Psi} + \frac{1}{H} \Delta \Psi - \frac{p_c}{2M_{Pl}^2 H}\Phi\;,
\label{jun14-2}\\
\Phi &= \frac{a}{a^\prime} \Psi^\prime = \frac{1}{H} \dot{\Psi} \; ,
\label{jun14-1}
\end{align}
where we specified to the model at hand by writing the second expressions.
We find from \eqref{jun14-1} that
\be
h_{00} =2 \Phi \sim \sqrt{\lambda} t^2 kv
\label{jun14-4}
\ee
and from \eqref{jun14-2} we see that $\d_i U = h_{0i}$ has two large
contributions
\be
h_{0i} \sim \sqrt{\lambda} t^2 k v
\label{jun14-5}
\ee
and
\be
h_{0i} \sim \sqrt{\lambda} t v \; .
\label{jun14-3}
\ee
The naive power counting suggests the UV scale $\Lambda^{(1)} $ in the
following way. There are cubic terms in the gravitational Lagrangian
\be
L_{int} = M_{Pl}^2 h \d h \d h \; .
\label{jun14-7}
\ee
With $h$ estimated by \eqref{jun14-3} this gives
\[
L_{int} = M_{Pl}^2 (\sqrt{\lambda} t)^3 v \d v \d v =
\frac{M_{Pl}^2}{\phi_c^3}  v \d v \d v \; .
\]
This can indeed be (mis)understood as the evidence for the UV scale
$\Lambda^{(1)}$ given by \eqref{4*}. Perturbations estimated according to
\eqref{jun14-4}, \eqref{jun14-5} naively yield strong coupling scale which
is even lower than $\Lambda^{(1)}$.

Likewise, the scale $\Lambda^{(2)}$  pops up in this gauge as follows. The
interaction term in the $\theta$ sector reads
\be
L_{int}^{h\chi \chi} = h \d \chi \d \chi\;,
\label{jun14-10}
\ee
where $\chi = \phi_c \theta$ is canonically normalized. The estimates
\eqref{jun14-4}, \eqref{jun14-5} give
\[
L_{int}^{h\chi \chi} = \sqrt{\lambda} t^2 \d v  \d \chi \d \chi \sim
\frac{1}{\sqrt{\lambda} \phi_c^2} \d v  \d \chi \d \chi\;.
\]
This can be viewed as an indication of the strong coupling scale
$\Lambda^{(2)} = \lambda^{1/4} \phi_c$.

\subsection{No strong coupling below $M_{Pl}$: Newtonian gauge}

Using the gauge $\varphi=0$ is, however, not appropriate for studying the
UV properties of the theory. In this gauge, metric perturbations are huge
at early times. The reason is that the field $\phi$ fluctuates, but
carries very little energy. Space-time is very close to Minkowskian, but
hypersurfaces of constant $\varphi$ are embedded in it in a  cumbersome
way.

Much clearer is the Newtonian gauge, in which $U=0$. In this gauge
\[
\Psi = - \Phi \; ,
\]
and $\Phi$ obeys
\be
\Delta \Phi = \frac{1}{2M_{Pl}^2} \phi_c^\prime \, \frac{z}{a}
\frac{\d}{\d \eta} \l \frac{v}{z} \r\;.
\label{jun29-3}
\ee
So, metric perturbation is small in this gauge at early times,
\be
\Phi \sim \frac{1}{M_{Pl}^2 \sqrt{\lambda} t^2} k^{-1} v \sim
\frac{\sqrt{\lambda} \phi_c^2}{M_{Pl}^2} k^{-1} v \; ,
\label{jun29-2}
\ee
where we again assume $k|t| \gg 1$. By plugging this into
\eqref{jun14-7} we see that no strong coupling scale is generated in UV
due to the interaction of $\varphi$ with metric perturbations. This shows
that there is actually no strong coupling in the $\phi$ sector of the
theory at momenta below $M_{Pl}$, i.e., the scale $\Lambda^{(1)}$ is
spurious.

The final point concerns the $\theta$ sector. The interaction term with
metric perturbation is given by \eqref{jun14-10}. In the Newtonian gauge
this gives
\[
L_{int}^{h\chi \chi} = \frac{1}{M_{Pl}^2 \sqrt{\lambda} t^2} \d^{-1} v
\d \chi \d \chi =  \frac{\sqrt{\lambda} \phi_c^2}{M_{Pl}^2} \d^{-1} v \d
\chi \d \chi \;.
\]
The dimensionless coupling here is small provided that the inequality
\eqref{jun28-3} holds. So, the scale $\Lambda^{(2)}$ is spurious too.

\section{Leading interactions of $\theta$}
\label{sec:non-linear}

As discussed in detail in
Refs.~\cite{Libanov:2010nk,Libanov:2010ci,Libanov:2011hh,Mironov-2} within
the spectator approximation, potentially interesting effects are due to
the interaction of the field $\theta$ with perturbations $\varphi$ at the
cubic level. In the dynamical  Hinterbichler--Khoury model, the relevant
terms in the action are of the zeroth and first order in perturbations of
$\phi$ and metric,
\[
S_\theta^{(0)} = \int~d^4x~ \frac{1}{2} a^2 \phi_c^2 \eta^{\mu \nu}
\d_\mu \theta \d_\nu \theta \; ,
\]
and
\be
S_\theta^{(1)} = \int~d^4x~ \frac{1}{2} a^2 \phi_c^2  \left[ \l -\Phi
+ 3 \Psi  + 2 \frac{\varphi}{\phi_c} \r \theta^{\prime \, 2} - \l \Phi +
\Psi  + 2 \frac{\varphi}{\phi_c} \r \d_i \theta \d_i \theta + 2 \d_i U
\theta^\prime \d_i \theta \right] \; ,
\label{jun25-5}
\ee
respectively, where we still partially fix the gauge according to
\eqref{jun29-1}. In the spectator model, one has $\Psi=\Phi=U=0$, while
the perturbation $\varphi$ coincides with $v$ in the short wavelength
regime $k|t| \gg 1$ and is given by~\cite{vr1}
\be
\varphi = - \frac{3}{k^2 t^2} v
\label{jun29-4}
\ee
in the large wavelength regime $k|t| \ll 1$. To see that the results
of the analysis in the spectator approximation are valid in the dynamical
model as well, it is again convenient to use the Newtonian gauge.

\subsection{Interactions in the Newtonian gauge}

In the first place, let us check that the perturbations $\varphi$ are the
same in the Newtonian gauge as in the spectator approximation. In the
short wavelength regime, one finds from \eqref{jun29-2} that
\[
\frac{a\phi_c^\prime}{a^\prime} \Psi \sim \frac{1}{kt} v \ll v \; ,
\]
so that $\varphi$ is indeed equal to $v$. In the large wavelength
regime Eq.~\eqref{jun29-3} gives
\be
\Phi = - \Psi =  \frac{1}{2M_{Pl}^2} \frac{\sqrt{2}}{\sqrt{\lambda}
t^3 k^2} v \; .
\label{jun29-5}
\ee
Therefore, we find from \eqref{jun28-6} that the leading part of
$\varphi$ is indeed given by \eqref{jun29-4}. As a cross check, it is
straightforward to see that the metric perturbations (given by
\eqref{jun29-2} and \eqref{jun29-5} in short and long wavelength limits,
respectively) give contributions to the field equation for $\varphi$ which
are suppressed at least by $(\lambda M_{Pl}^2 t^2)^{-1}$. Hence, gravity
does not modify $\varphi$ in the Newtonian gauge.

Now, in the Newtonian gauge one always has
\[
\Phi \ll \frac{\varphi}{\phi_c} \; .
\]
Indeed, it follows from \eqref{jun29-2} that in the short wavelength
limit the gravitational potential is doubly suppressed,
\[
\frac{\Phi}{\varphi/\phi_c} = \frac{\Phi \phi_c}{v} \sim
\frac{1}{\lambda M_{Pl}^2 t^2} \frac{1}{kt} \;,
\]
while it follows from \eqref{jun29-4} and \eqref{jun29-5} that in the
large wavelength limit
\[
\frac{\Phi}{\varphi/\phi_c} \sim \frac{1}{\lambda M_{Pl}^2 t^2} \; .
\]
Hence, the gravitational potentials are subdominant in the action
\eqref{jun25-5}, as compared to $\varphi$. This shows that calculations in
the spectator approximation give correct results in the dynamical model,
modulo corrections suppressed by $(\lambda M_{Pl}^2 t^2)^{-1}$.

\subsection{Statistical anisotropy in $\varphi=0$ gauge}

The statistical anisotropy in the power spectrum of $\theta$, and hence in
the resulting adiabatic perturbations, is due to the interaction of the
field $\theta$ with those modes of $\varphi$ and metric which are still
superhorizon today. The analysis of this effect has been performed in
Ref.~\cite{Libanov:2010nk} in the spectator approximation. Here we repeat
this analysis in the gauge $\varphi=0$ to see that the results of the
spectator approximation are indeed reproduced, and also get better
understanding of the cancellation of infrared effects.

It will become clear shortly that as long as $\phi_c \ll M_{Pl}$, our
analysis is valid for both the spectator model of Ref.~\cite{vr1} and the
dynamical Hinterbichler--Khoury model. For the time being, we specify to
the dynamical scenario.

We are interested in the propagation of the field $\theta$ in perturbed
background in the case when the wavelength of $\theta$ is much shorter
than the wavelength of $\varphi$ and metric perturbations. Indeed, the
relevant adiabatic perturbations, and hence the modes of $\theta$, are
subhorizon today, while the perturbations of $\varphi$ and metric,
responsible for the statistical anisotropy, are still superhorizon. So, we
consider $\varphi$ and metric perturbations in the large wavelength
regime. In the gauge $\varphi = 0$ we find from \eqref{jun28-6} that
\[
\Psi = - \frac{1}{3\sqrt{2} \sqrt{\lambda} t M_{Pl}^2} v \; ,
\]
while the constraint equations \eqref{jun14-2} and \eqref{jun14-1} give
\begin{align}
\Phi & =  \frac{\sqrt{\lambda}}{\sqrt{2}} t v\;, \\
\Delta U &= - \frac{3}{\sqrt{2}} \sqrt{\lambda} v \; ,
\label{jun30-1}
\end{align}
where we made use of the fact that the terms with $\Psi$ in
\eqref{jun14-2} are small compared to the last term. The result
\eqref{jun30-1} can be understood as follows. In the Newtonian gauge (as
well as in the spectator approximation), large wavelength perturbation
\eqref{jun29-4} of the field $\phi$ can be viewed as inhomogeneous
time-shift of the background field \eqref{1**} (cf. Refs.~\cite{vr1,Cr}),
\[
\delta t ({\bf x})= - \frac{3 \sqrt{\lambda}}{\sqrt{2}}
\frac{1}{\Delta} v ({\bf x}) \; .
\]
Gauge transformation to the gauge $\varphi = 0$ corresponds to the
change of the time coordinate $t \to t + \delta t$. This induces
\be
U ({\bf x}) = \delta t ({\bf x}) \; ,
\label{jun30-2}
\ee
which is precisely the result \eqref{jun30-1}. Note that since $v$ is
a free canonically normalized massless field in effectively Minkowski
space-time, the power spectra of both $\varphi$  and $\delta t$ in the
Newtonian gauge are red, but the power spectrum of metric perturbation
$h_{0i} = \d_i U$ is flat in the gauge $\varphi=0$, while  $\Phi$ and
$\Psi$ have blue power spectra in that gauge. This explains why the power
law infrared enhancement of $\varphi$, visible in the spectator
approximation, is actually irrelevant. Note also that Eq.~\eqref{jun30-2}
shows that the form of $U$ is  common to the dynamical and spectator
models, so the analysis below and the result \eqref{jul4-1} are valid in
both of these models.

Non-vanishing $U({\bf x})$ yields the anisotropic last term in the action
\eqref{jun25-5}, We are going to keep at most two derivatives of $U$, and
study anisotropy proportional to $\d_i \d_j U$: since we are interested in
very large wavelengths of the perturbed background, higher derivatives are
suppressed. For the canonically normalized field $\chi = \phi_c \theta$ we
get, omitting isotropic terms,
\[
S_\theta^{(1)} = \int~d^4x~  2 \d_iU \dot{\chi} \d_i \chi\;.
\]
Therefore, the field equation reads, again modulo isotropic
corrections,
\be
\ddot \chi - \Delta \chi - \frac{2}{t^2} \chi + 2 \d_i U \d_i
\dot{\chi} = 0 \; .
\label{jun25-6}
\ee
Let us search for the solution in the following form:
\be
\chi = \mbox{e}^{i{\bf kx} - ikt -ik U({\bf x})} \l 1 - \frac{i}{(k -
k_i \d_i U) t} \r +  \mbox{e}^{i {\bf kx} -ik U({\bf x})} F^{(2)} (k, t)
\; ,
\label{jun25-7}
\ee
where $F^{(2)} \propto \d_i \d_j U$ and $F^{(2)}$ tends to zero as $t
\to -\infty$. Note that in the asymptotic past, this solution is a plane
wave, $\chi = \mbox{e}^{i{\bf kx} - ikt}$, modulo the time-shift
\eqref{jun30-2}. By substituting the Ansatz \eqref {jun25-7} into
Eq.~\eqref{jun25-6}, we obtain the equation for $F^{(2)}$:
\[
\ddot{F}^{(2)} +k^2  F^{(2)} - \frac{2}{t^2} F^{(2)} = 2 k_i k_j \d_i
\d_j U \mbox{e}^{-ikt} \frac{1}{k^2 t} \; .
\]
Up to notations, this equation coincides with the anisotropic part of
Eq.~(4.5) of Ref.~\cite{Libanov:2010nk}. As $t \to 0$, the imaginary part
of its solution is
\[
\mbox{Im}~F^{(2)} = \frac{\pi}{2} \frac{1}{k^2 t} \cdot
\frac{k_ik_j}{k^2} \d_i \d_j U \; .
\]
Thus, in the large wavelength regime the mode function of the field
$\theta = \chi/\phi_c$ is
\be
\theta = -  \frac{i}{q} \mbox{e}^{i{\bf qx}} \l 1 - \frac{\pi}{2k}
\cdot \frac{k_ik_j}{k^2} \d_i \d_j U \r \; ,
\label{jun30-10}
\ee
where $q_i = k_i - \d_i U$, and we omitted irrelevant isotropic
corrections, as well as real part of $F^{(2)}$. The latter does not
contribute to the statistical anisotropy to the linear order in $U$ (and
hence in $\sqrt{\lambda}$), which comes from the interference of the two
terms in parentheses in \eqref{jun30-10}. So, the power spectrum of
$\theta$, and hence of the adiabatic perturbations $\zeta$, contains
anisotropic part,
\be
{\cal P}_\zeta = {\cal P}^{(0)}_\zeta \l 1 - \frac{\pi}{k} \cdot
\frac{k_ik_j}{k^2} \d_i \d_j U \r \; ,
\label{jul4-1}
\ee
where  ${\cal P}^{(0)}_\zeta$ is isotropic flat power spectrum. This
is the result of Ref.~\cite{Libanov:2010nk}. Since we consider modes of
$U$ which are still superhorizon today, the tensor $\d_i \d_j U$ is
constant throughout the visible Universe, and accounting for modes of $U$
with the momenta below $H_0$ only, we obtain from Eq.~\eqref{jun30-1}  the
estimate for its strength, $\d_i \d_j U \sim \sqrt{\lambda}  H_0 $.

\section{Conclusions}
\label{sec:conclusion}

To summarize, in this note we have made two simple observations. First, we
have argued that mixing of scalar field(s) with metric in dynamical
(pseudo-)conformal models does not introduce new UV strong coupling
scales. Although we have explicitly considered the Hinterbichler--Khoury
model, this conclusion appears generic. As another example, metric
perturbations do decouple~\cite{Cr} in the Galilean Genesis model in the
Newtonian gauge in the limit $M_{Pl} \to \infty$. A phenomenological
consequence of our first observation is that one does not have to impose
strong constraints on the parameters of  dynamical models to ensure
self-consistency; the self-coupling $\lambda$  of the
Hinterbichler--Khoury model (and its analog $\Lambda^3/f^3$ in the
Galilean Genesis, see Ref.~\cite{Mironov-2}) need not be particularly
small. Since these parameters govern the non-linear effects, such as
statistical anisotropy and non-Gaussianity, we conclude that these effects
may have observable strengths.

Our second observation is that the spectator approximation does give
correct results in dynamical models, provided the background space-time is
sufficiently flat (which is likely to be a pre-requisite for the flat
scalar spectrum in dynamical conformal models). A qualification is that
this applies to the field $\theta$ and its interactions with perturbations
of $\phi$ and metric: curvature perturbations generated directly by
perturbations of the rolling field $\phi$ and their mixing with metric are
entirely different story~\cite{Cr,HK} --- and they are negligible anyway.
A consequence of our second observation is that potentially observable
effects discussed in
Refs.~\cite{Libanov:2010nk,Libanov:2010ci,Libanov:2011hh,Mironov-2} are
inherent in the entire class of both spectator and dynamical conformal
models.

\section*{Acknowledgements}
The authors are indebted to S.~Mironov and S.~Ramazanov for useful
comments and discussions. We thank  K.~Hinterbichler and J.~Khoury for
stimulating correspondence. This work has been supported in part by the
Federal Agency for Science and Innovations under state contract
02.740.11.0244 and by the grant of the President of the Russian Federation
NS-5525.2010.2. The work of M.L. has been supported in part by Russian
Foundation for Basic Research grant 11-02-92108 and by the Dynasty
Foundation. The work of V.R. has been supported in part by the SCOPES
program.


\begin{thebibliography}{99}
\bibitem{vr1}
  V.~A.~Rubakov,
  ``Harrison--Zeldovich spectrum from conformal invariance'',
  {JCAP} {\bf 0909} (2009), 030;
   arXiv:0906.3693 [hep-th].


\bibitem{Cr}
 P.~Creminelli, A.~Nicolis and E.~Trincherini,
  ``Galilean Genesis: an alternative to inflation,''
  JCAP {\bf 1011}, 021 (2010);
  arXiv:1007.0027 [hep-th].

\bibitem{HK}
  K.~Hinterbichler and J.~Khoury,
  ``The Pseudo-Conformal Universe: Scale Invariance from Spontaneous Breaking
  of Conformal Symmetry,''
  arXiv:1106.1428 [hep-th].


\bibitem{Antoniadis:1996dj}
  I.~Antoniadis, P.~O.~Mazur and E.~Mottola,
  ``Conformal invariance and cosmic background radiation,''
  Phys.\ Rev.\ Lett.\  {\bf 79} (1997) 14
  [arXiv:astro-ph/9611208].

\bibitem{Linde:1996gt}
  A.~D.~Linde and V.~F.~Mukhanov,
  {Nongaussian isocurvature perturbations from inflation},
  {Phys.\ Rev.}\  D {\bf 56} (1997), 535
   [astro-ph/9610219];\\
  K.~Enqvist and M.~S.~Sloth,
  ``Adiabatic CMB perturbations in pre big bang string cosmology'',
  {Nucl.\ Phys.}\  B {\bf 626} (2002), 395
   [hep-ph/0109214];\\
  D.~H.~Lyth and D.~Wands,
  ``Generating the curvature perturbation without an inflaton'',
  {Phys.\ Lett.}\  B {\bf 524} (2002), 5
   [hep-ph/0110002];\\
  T.~Moroi and T.~Takahashi,
  ``Effects of cosmological moduli fields on cosmic microwave background'',
  {Phys.\ Lett.}\  B {\bf 522} (2001), 215
  [{Erratum-ibid.}\  B {\bf 539} (2002), 303]
   [hep-ph/0110096].

\bibitem{Dvali:2003em}
  G.~Dvali, A.~Gruzinov and M.~Zaldarriaga,
``A new mechanism for generating density perturbations from inflation'',
  {Phys.\ Rev.}\  D {\bf 69} (2004), 023505
   [astro-ph/0303591];\\
  L.~Kofman,
 ``Probing string theory with modulated cosmological fluctuations'',
  astro-ph/0303614;\\
  G.~Dvali, A.~Gruzinov and M.~Zaldarriaga,
  ``Cosmological perturbations from inhomogeneous reheating, freezeout, and
    mass domination'',
  {Phys.\ Rev.}\  D {\bf 69} (2004), 083505
   [astro-ph/0305548].


\bibitem{Libanov:2010nk}
  M.~Libanov and V.~Rubakov,
  ``Cosmological density perturbations from conformal scalar field: infrared
  properties and statistical anisotropy,''
  JCAP {\bf 1011} (2010) 045
  [arXiv:1007.4949].

\bibitem{Libanov:2010ci}
  M.~Libanov, S.~Mironov and V.~Rubakov,
  ``Properties of scalar perturbations generated by conformal scalar field,''
  arXiv:1012.5737 [hep-th].


\bibitem{Libanov:2011hh}
  M.~Libanov, S.~Ramazanov and V.~Rubakov,
  ``Scalar perturbations in conformal rolling scenario with intermediate
  stage,''
  JCAP {\bf 1106} (2011) 010
  [arXiv:1102.1390 [hep-th]].

\bibitem{Mironov-2}
 M.~Libanov, S.~Mironov and V.~Rubakov,
  ``Non-Gaussianity of scalar perturbations generated by conformal
  mechanisms,''
  arXiv:1105.6230 [astro-ph.CO].

\bibitem{ekpyro-i}
  J.~Khoury, B.~A.~Ovrut, P.~J.~Steinhardt and N.~Turok,
  ``The ekpyrotic universe: Colliding branes and the origin of the hot big
  bang,''
  Phys.\ Rev.\  D {\bf 64} (2001) 123522
  [arXiv:hep-th/0103239];\\
J.~Khoury, B.~A.~Ovrut, N.~Seiberg, P.~J.~Steinhardt and N.~Turok,
  ``From big crunch to big bang,''
  Phys.\ Rev.\  D {\bf 65} (2002) 086007
  [arXiv:hep-th/0108187].

\bibitem{Nicolis:2008in}
  A.~Nicolis, R.~Rattazzi and E.~Trincherini,
  ``The Galileon as a local modification of gravity,''
  Phys.\ Rev.\  D {\bf 79} (2009) 064036
  [arXiv:0811.2197].

\bibitem{Levasseur:2011mw}
  L.~P.~Levasseur, R.~Brandenberger and A.~C.~Davis,
  ``Defrosting in an Emergent Galileon Cosmology,''
  arXiv:1105.5649 [astro-ph.CO] .

\bibitem{Gao:2011mz}
  X.~Gao,
  ``Conserved cosmological perturbation in Galileon models,''
  arXiv:1106.0292 [astro-ph.CO].

\end{thebibliography}
\end{document}